\documentclass[aps,twocolumn,pra,showpacs,superscriptaddress,10pt]{revtex4-1}
\usepackage{mathtools}
\usepackage{amssymb}
\usepackage{color}
\usepackage{graphicx}
\usepackage{epsfig}
\usepackage{epstopdf}
\usepackage{hyperref}
\usepackage{dsfont}
\usepackage{todonotes}
\usepackage{physics}

\begin{document}

\title{Polarization selective Dove prism}
\author{P.~A.~Ameen Yasir}
\email{ameenyasir.p.a@gmail.com}
\affiliation{Department of Physical Sciences, Indian Institute of Science Education and research, Mohali, Punjab 140306, India}
\author{Sandeep K.~Goyal}
\email{skgoyal@iisermohali.ac.in}
\affiliation{Department of Physical Sciences, Indian Institute of Science Education and research, Mohali, Punjab 140306, India}

\begin{abstract}
  We propose a passive all optical device capable of transforming the orbital angular momentum (OAM) state of light conditioned over the polarization states. The efficiency of this device is ensured due to its linear optical nature. As applications of this device, we show CNOT and SWAP operations between polarization and OAM qubits, non-interferometric  OAM mode sorter and generalized Pauli $X$ operation on a four-dimensional subspace of OAM. 
\end{abstract}

\maketitle

\section{Introduction} \label{s1}
Quantum information processing (QIP) which include quantum communication and quantum computation is the fastest going field in science and technology. QIP exploits the superposition and measurement collapse of the  quantum systems to gain higher speed up for computational purposes and to make communication unconditionally secure~\cite{Steane1998,nielsen2010}. Although, quantum computation can be performed using a large variety of systems such as trapped  atom and ions~\cite{Lukin2001,HAFFNER2008,Benhelm2008,Preskill2018},  superconducting circuits~\cite{Kawabata2004,Gambetta2017,Huang2020}, Nuclear magnetic resonance~\cite{Xin2018,Jones2001}, defect centers in diamonds~\cite{Weber2010,Wu2019,Nizovtsev2005}, and photonic systems~\cite{OBrien2007,Barz2015,Kok2007,Wang2016,Flamini2019}, photons are the only viable option for long distance quantum communication because of their weak interaction with their environment.
Photons possess several degrees of freedom (DoFs) such as polarization, frequency, and orbital angular momentum (OAM) which can be used for quantum communication tasks~\cite{Flamini2019}. 

OAM states are the eigenmodes of the paraxial wave equation described by helical wavefronts, which is characterized by the winding number $-\infty < \ell < \infty$.  OAM of light can serve as a multi-level quantum system as it possess infinite number of orthogonal states~\cite{Erhard2018}. However, implementing a general unitary operation on these states can be resource intensive~\cite{Escartin2011}. Furthermore, the difficulty in manipulating and measuring the states of OAM restricts its usability.

An efficient and general unitary operation is known  only on the subspace spanned by $\ell = \pm 1$~\cite{padgett99b}. Similarly, no efficient method to perform measurement on the OAM space is known~\cite{yao2011,gbur2017}. The most common methods to measure the OAM states of light involves cumbersome interferometric setups in order to sort orthogonal  OAM modes in different spacial modes~\cite{leach2004}, or performing log-polar transformation~\cite{bryngdahl74} on them to convert the helical phase structure into linear phase and sort them through the use of a convex lens~\cite{berkhout2010}. Other methods which make use of log-polar transformation -- directly or indirectly -- to sort LG modes include~\cite{mirhosseini2013,sahu2018,lightman2017}.

OAM along with polarization DoF of light can also be used as a hybrid quantum system for QIP tasks. However, coupling the OAM and polarization is a challenge on its own. Devices such as $q$-plate~\cite{marrucci2006} and $J$-plate~\cite{devlin2017} can, for instance, provide this coupling. However, generation of vector-vortex beams using such devices have been only demonstrated for two superpositions~\cite{ambrosio2016,mclaren2015}, and generation of higher-order superpositions remains a difficult process. Moreover, fabrication of such devices can be fairly difficult.

Here we propose an all optical device to couple the polarization and OAM DoFs. The action of this device can be described as the action of Dove prism (DP) on the OAM DoF conditioned over the polarization states of light; hence, we call it polarization sensitive dove prism (PSDP). It consists of two half-wave-plates and a cube formed by gluing three negative uniaxial crystals together in a specific manner. We discuss a few  application of PSDP such as non-interferometric OAM mode sorter, CNOT and SWAP operations between OAM and polarization states,  and generalized four-dimensional Pauli $X$ operation on the OAM modes. PSDP is a passive, all optical device which consists of lossless linear optical elements, which makes this device  efficient and scalable, suitable for optical quantum computation and quantum communication.

The article is organized as follows: in Sec.~\ref{s2} we present the relevant background required to understand our result. Here we discuss vector-vortex beams, Dove prism, techniques used to sort  OAM modes, and generalized Pauli $X$ operation. In Sec.~\ref{s3} we present the details of PSDP. The applications of PSDP are presented in Sec.~\ref{s4}. We conclude in Sec.~\ref{s6}.

\section{Background} \label{s2}
In this section, we introduce the concepts relevant for our results. We start with vector-vortex beams -- beams whose OAM and polarization DoF are non-separable. We describe the DP and methods to sort OAM states. Finally, we end the section with describing generalized Pauli $X$ operation on OAM DoF. 

\subsection{Vector-vortex beams}
Orbital angular momentum states are the solutions of the paraxial wave equation in the polar coordinates $(r,\phi,z)$ and are represented by Laguerre-Gaussian (LG) modes~\cite{kogelnik66,allen92}. These OAM modes can be characterized by two indices, $(\ell, p)$. While azimuthal index $\ell$ can assume any  integer value  $\{\ldots,-1,0,1,\ldots\}$, the radial index $p$ can take values $\{0,1,2,\ldots\}$. LG mode $\psi_p^{\ell} (r,\phi;z)$ is also an eigenmode of the angular momentum operator $-i\hbar\partial/\partial \phi$; hence, they carry OAM of magnitude $\ell\hbar$ per photon~\citep{allen92}. To this end, we can write~\cite{sakurai94} 
\begin{align} \label{oam3}
\psi_p^{\ell} (r,\phi;z) = \langle x,y|\psi_p^{\ell}(z) \rangle,
\end{align}
i.e., $\psi_p^{\ell} (r,\phi;z)$ is the position representation of the vector $|\psi_p^{\ell}(z) \rangle$.

The state vectors $\{|\psi_p^{\ell}(z) \rangle\}$ form a complete orthonormal basis for the Hilbert space $\mathcal{H}$ of the transverse modes of paraxial light~\cite{simon93}. Since, the transverse modes are independent of the polarization, an arbitrary state vector for the polarization and the transverse modes can be written as~\cite{forbes2014}
\begin{align} \label{oam4}
|\Psi \rangle = \sum_{\ell,p} c_{\ell,p} |\psi_p^{\ell}(z) \rangle \otimes |h \rangle + \sum_{\ell',p'} c'_{\ell',p'} |\psi_{p'}^{\ell'}(z) \rangle \otimes |v \rangle,
\end{align} 
where $c_{\ell,p}$ and $c'_{\ell',p'}$ are complex coefficients. It can happen that the mode $|\Psi \rangle$ is entangled in the spatial and polarization modes of freedom. Such modes are said to be ``classically entangled''~\cite{spreeuw98} and are known as vector-vortex beams~\cite{forbes2014,mclaren2015,ambrosio2016}. 

\subsection{Dove prism} \label{dov}

\begin{figure}
\centering
\includegraphics[scale=0.4]{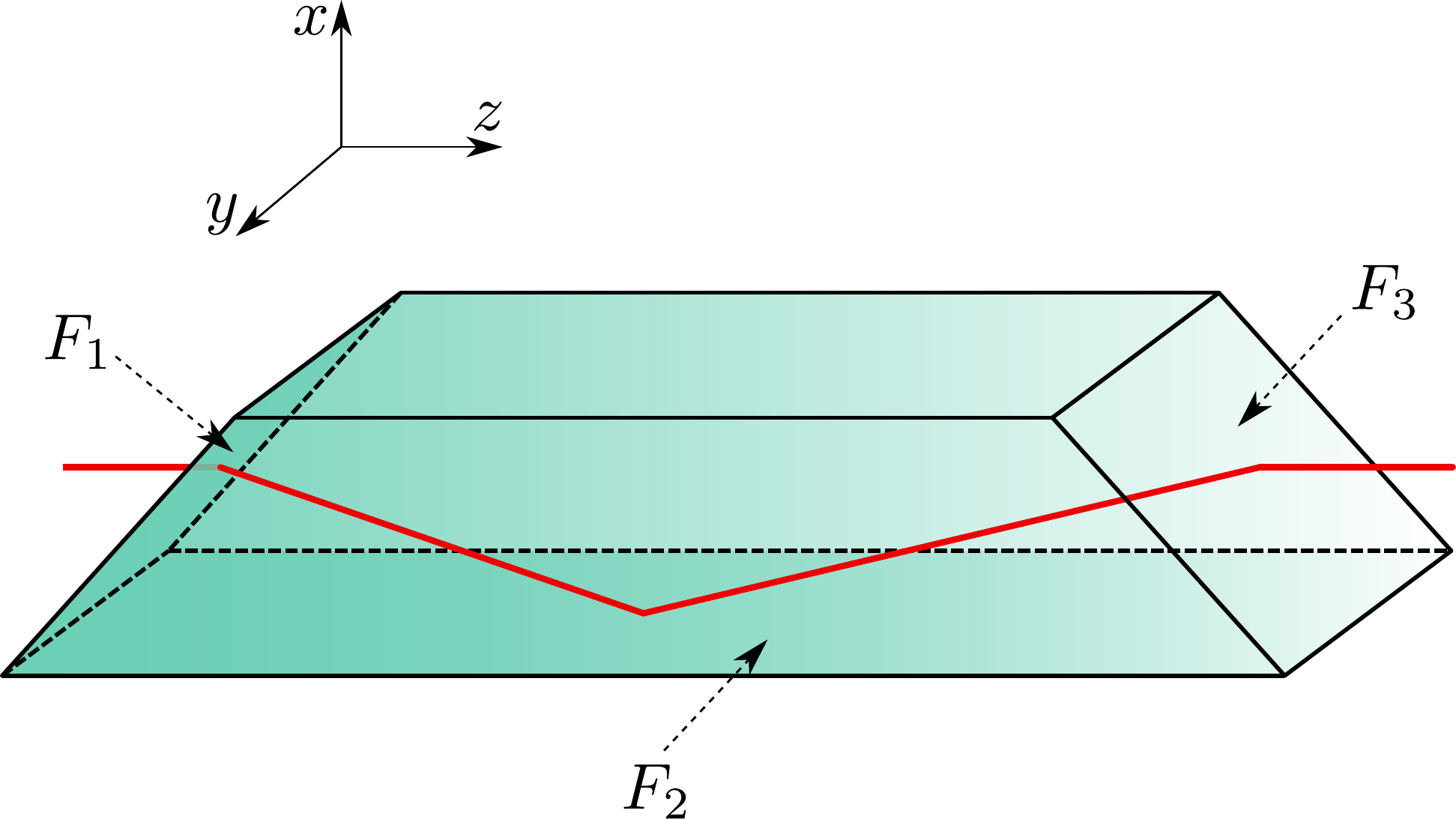}
\caption{Shown here is schematic of the DP placed in $y$-$z$ plane. Paraxial light field $\psi(x,y;z)$ traveling in the $z$-direction first gets refracted by face $F_1$, total internally reflected by face $F_2$, and finally gets refracted by face $F_3$. The DP transforms the paraxial field as $\psi(x,y;z) \rightarrow \psi(-x,y;z)$\,[see (\ref{dov1})].}
\label{dove}
\end{figure}

 A DP is a reflective type prism which flips the image in one transverse direction and leaves the other unchanged. Suppose a DP is placed along the $z$-axis (as shown in Fig.\,\ref{dove}) and a paraxial light field $\psi(x,y;z)$ propagating in the $z$-direction passes through it. The action of the DP on this light field can be written as
\begin{align} \label{dov1}
\psi(x,y;z) \rightarrow \psi(-x,y;z).
\end{align} 
In other words, $r \rightarrow r$ while $\phi \rightarrow -\phi$. As a result, the LG mode $\psi_p^{\ell} (r,\phi;z)$ transforms as
\begin{align} \label{dov2}
  \psi_p^{\ell} (r,\phi;z) \rightarrow \psi_p^{-\ell} (r,\phi;z).
\end{align}
If we represent the DP action by the operator $\mathcal{U}_s$ where the subscript $s$ denotes that the operator $\mathcal{U}$ acts only on the spatial modes, then we can write 
\begin{align} \label{dov2a}
\mathcal{U}_s \ket{\psi^\ell_p} = \ket{\psi^{-\ell}_p}.
\end{align}
If the DP is rotated through an angle $\alpha$ about the $z$-axis anticlockwise, then the operator is represented by $\mathcal{U}_s(\alpha)$ and the LG mode $\ket{\psi_p^{\ell}}$ transforms as~\cite{gonzalez2006}
\begin{align} \label{dov3}
\mathcal{U}_s(\alpha)\ket{\psi_p^{\ell}} = e^{\text{i} 2\ell\alpha} \ket{\psi_p^{-\ell}}.
\end{align}
The reflection suffered by the light beam can result in the change of polarization. However, DPs can be designed such that the polarization of the light remains unaffected~\cite{padgett99}, also known as ``idealized DP''~\cite{leach2004}.

\subsection{OAM sorting} \label{oas}

\begin{figure}
\centering
\includegraphics[scale=0.4]{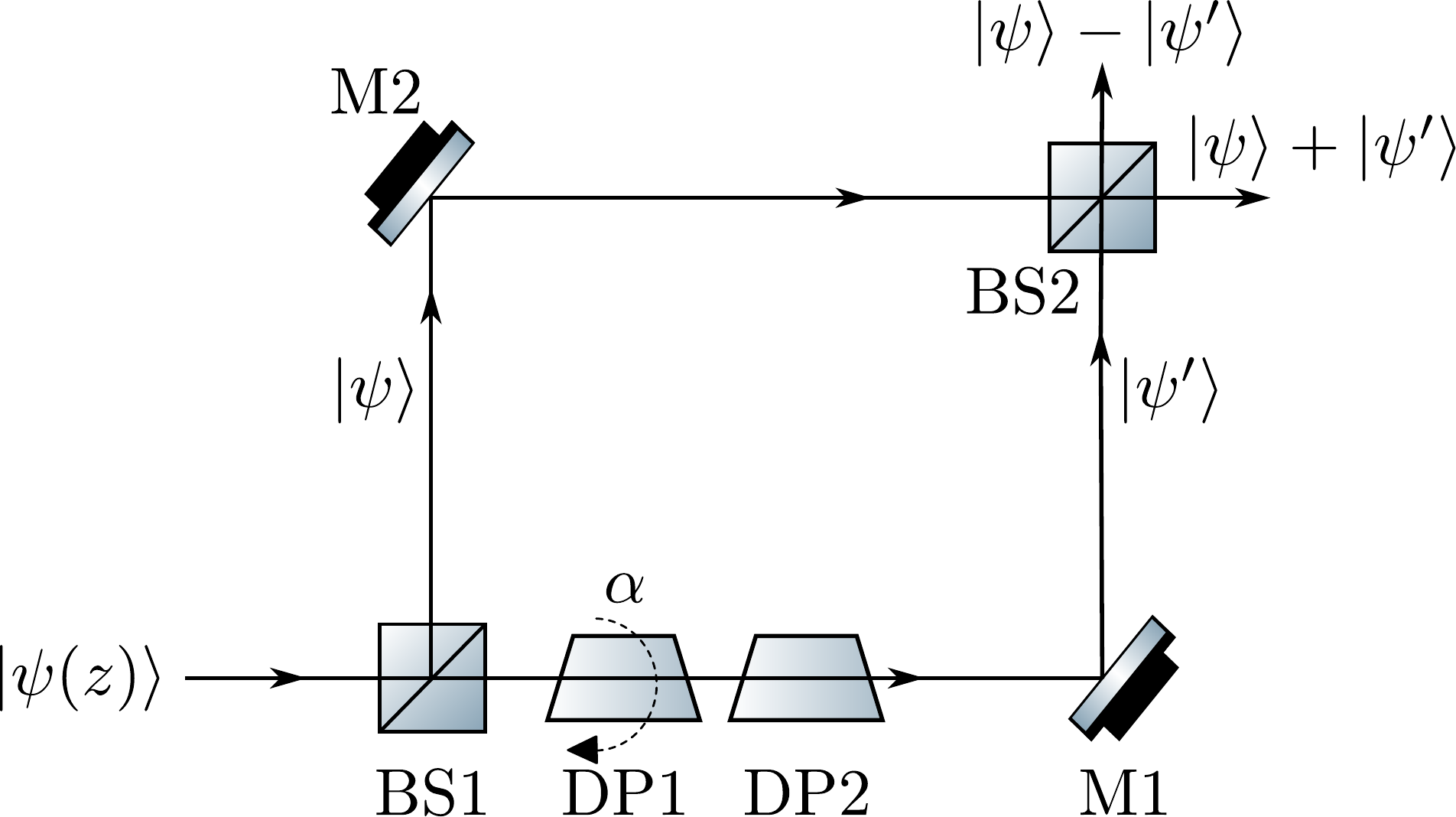}
\caption{Mach-Zehnder\,(MZ) interferometric setup to sort linear combination of $N$ LG modes as given in Eq.\,(\ref{pad1}). Balanced beamsplitter BS1 splits the input paraxial light field $|\psi \rangle$ into two fields which are both proportional to $|\psi \rangle$. While one of them reaches beamsplitter BS2 after getting reflected by mirror M2, the other passes through 2 DPs\,(where the first one is rotated through an angle $\alpha$ about the $z$-axis with respect to the second), reflected by mirror M1, and reaches BS2. Interference of these two light fields, $|\psi \rangle$ and $|\psi' \rangle$\,[see (\ref{pad2})], at BS2 results in the light fields $(|\psi \rangle + |\psi' \rangle)/\sqrt{2}$ and $(|\psi \rangle - |\psi' \rangle)/\sqrt{2}$ at the output ports. Choosing the angle of rotation $\alpha$ to be $\pi/2$ yields the even and odd OAM modes on different output ports.}
\label{mz1}
\end{figure}

Here, we briefly review some of the methods available in the literature to sort the LG modes present in a paraxial light beam and thereby measure the OAM of the paraxial light field.  We assume that the paraxial light field $\psi(r,\phi;z)$ contains $N$ number of LG modes with $p=0$ alone, and let $|\psi_{0}^{\ell}(z) \rangle \equiv |\ell \rangle$. With this we can write the vector $|\psi(z) \rangle$\,(corresponding to the paraxial field $\psi(x,y;z)$) as
\begin{align} \label{pad1}
|\psi(z) \rangle = \sum_{\ell=0}^{N-1} c_\ell \,|\ell \rangle,
\end{align}
where $c_\ell$'s are complex coefficients.

We first explain the interferometric method proposed in Ref.~\cite{leach2004}. In this method, a setup that consists of a Mach-Zehnder (MZ) interferometer embedded with two DPs is used in order to sort even and odd OAM modes~(Fig.~\ref{mz1}). Sending $\ket{\psi(z)}$ through the first beamsplitter BS1 in the MZ setup in Fig.\,\ref{mz1}, we see that one of the light fields is reflected by mirror M2 and reaches beamsplitter BS2. The other light field passes through two DPs, reflected by mirror M1, and reaches beamsplitter BS2. These two DPs are kept such that the first one is rotated through an angle $\alpha$ about the $z$--axis with respect to the second one. So the light field passing through the DPs, say $|\psi'(z) \rangle$, transforms according to (\ref{dov3}) as
\begin{align} \label{pad2}
|\psi(z) \rangle \rightarrow |\psi'(z) \rangle = \mathcal{U}_s\mathcal{U}_s(\alpha)\ket{\psi_p^\ell} = \sum_{\ell=0}^{N-1} c_\ell\,e^{\text{i} 2\ell\alpha} \,|\ell \rangle.
\end{align}
On interfering the two light fields $|\psi(z) \rangle$ and $|\psi'(z) \rangle$ on BS2, we obtain $\frac{1}{\sqrt{2}}(|\psi(z) \rangle +|\psi'(z) \rangle)$ and $\frac{1}{\sqrt{2}}(|\psi(z) \rangle -|\psi'(z) \rangle)$ respectively.  For the choice $\alpha=\pi/2$, it is readily seen that even\,($|0 \rangle$, $|2 \rangle$, $|4 \rangle$, $\ldots$) and odd\,($|1 \rangle$, $|3 \rangle$, $|5 \rangle$, $\ldots$) LG modes are sorted out and are available at the two output ports of BS2.

\begin{figure}
\centering
\includegraphics[scale=0.25]{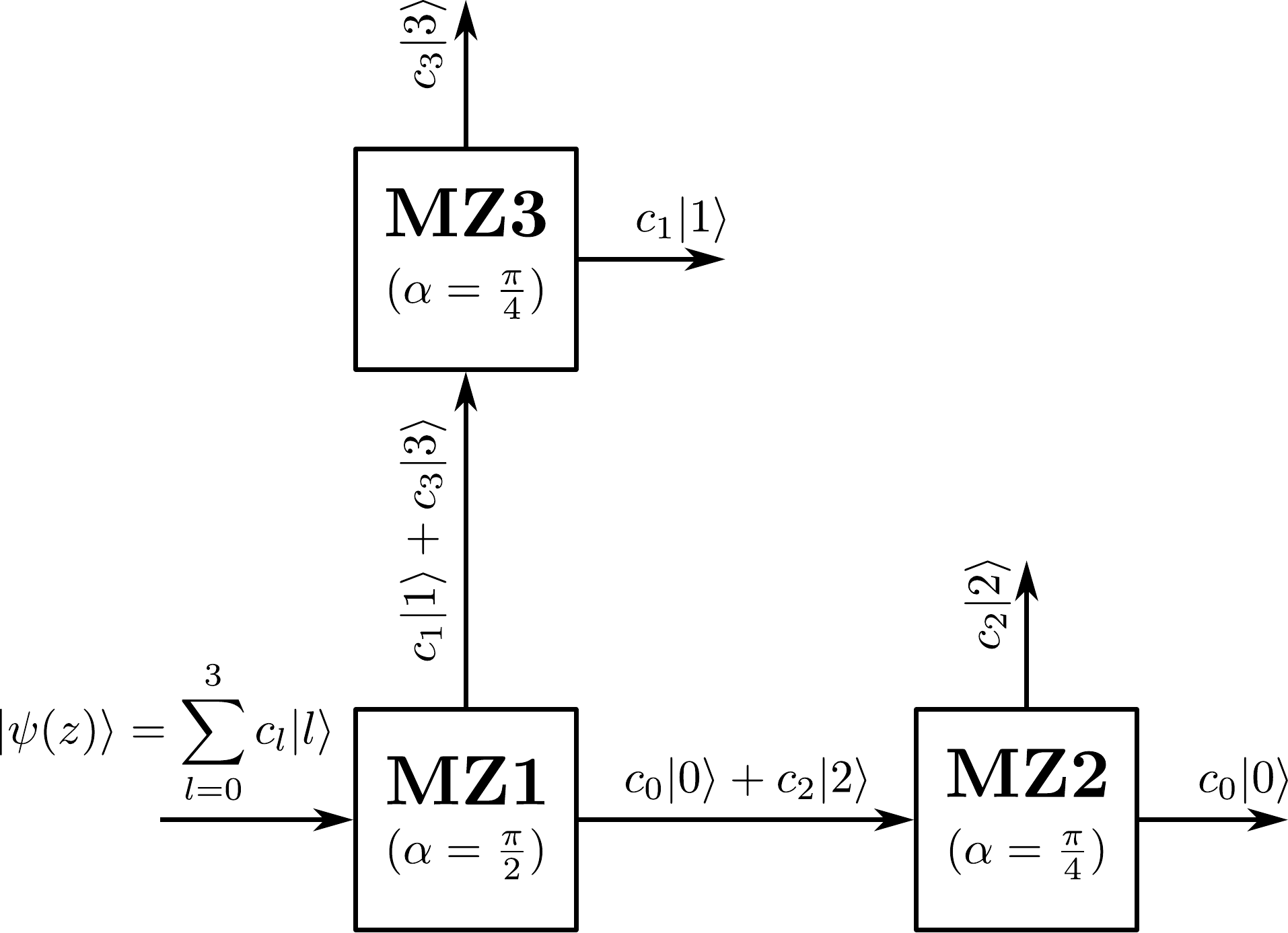}
\caption{Demonstrating how a linear combination of four LG modes is sorted out using three MZ interferometers described in the Fig.~\ref{mz1}. MZ1\,(with $\alpha= \pi/2$) sorts even and odd OAM modes. While the LG modes $|0 \rangle$ and $|2 \rangle$ are sorted by MZ2\,(with $\alpha=\pi/4$), $|1 \rangle$ and $|3 \rangle$ are sorted by MZ3\,(with $\alpha=\pi/4$).}
\label{mz2}
\end{figure}

Now the even LG modes can be sorted out using a similar MZ setup with the choice $\alpha=\pi/4$ as two sets of LG modes $\{|0 \rangle, |4 \rangle, |8 \rangle, \ldots\}$ and $\{|2 \rangle, |6 \rangle, |10 \rangle, \ldots\}$, respectively. Meanwhile, the odd LG modes can be sorted out using another MZ setup with the choice $\alpha=\pi/4$ as two sets of LG modes $\{|1 \rangle, |5 \rangle, |9 \rangle, \ldots\}$ and $\{|3 \rangle, |7 \rangle, |11 \rangle, \ldots\}$, respectively. Hence, recursively one can sort all $N$ individual LG modes using $N-1$ MZ interferometers with appropriate choice of $\alpha$. Sorting of 4 LG modes using 3 MZ interferometers has been illustrated in Fig.~\ref{mz2}. Even though this is a simple method to sort OAM states, alignment of $N-1$ interferometers is a difficult task, which restricts the scalability of this method.

Another approach to sort OAM modes is based on Cartesian to log-polar coordinate transformation which was proposed and experimentally demonstrated by Berkhout\,{\it et\,al}.~\cite{berkhout2010}. This optical transformation maps the azimuthal phase profile pertaining to an LG mode to a tilted planar wavefront. Now a convex lens focuses individual LG modes to a different position and thereby sorting the LG modes. However, the separation efficiency was just 77\%. By introducing refractive beam-copying device\,(implemented using spatial light modulator) along with log-polar optical transformation Mirhosseini\,{\it et\,al}.~\cite{mirhosseini2013} reported a separation efficiency of 92\% experimentally. 

\subsection{Generalized Pauli {\it X} operation} \label{gs}

Generalized Pauli $X$ operation and its integer  powers along with the generalized Pauli $Z$ (with its integer powers) can be used to implement an arbitrary unitary operation on a given dimensional Hilbert space. The $Z$ operation\,(in $N$ dimensions) given by $Z = \sum_{n=0}^{N-1}\exp{2i\pi n/N}\ket{n}\bra{n}$ can be easily implemented on OAM space by using two DPs, where the first one is rotated through an angle $\alpha=\pi n/N$ about the $z$-axis with respect to the second one~\cite{palici2020}. However, implementing $X$ is not as simple. Four-dimensional implementation of $X$ on OAM space can be achieved as given below~\cite{Babazadeh2017,Gao2019}.

Consider the OAM subspace spanned by the following four spatial modes $\{\ket{-2},\ket{-1}, \ket{0}, \ket{1} \}$. The action of $X$ can be written as 
\begin{align} \label{gs1}
X \ket{\ell} = \ket{\ell\oplus 1},
\end{align}
where $\ell \in \{-2,-1,0,1\}$ and $\ell \oplus 1=\ell+1\,\,({\rm mod}\,\,4)$. In other words, $X$ is a cyclic permutation operation.

The $X$ operation can be realized as follows~\cite{schlederer2016}. First, the transformation $\ket{\ell} \rightarrow \ket{\ell+1}$ is achieved using a spiral phase plate, which adds an OAM of $\hbar$ per photon. With this, we obtain $\{\ket{-2},\ket{-1}, \ket{0}, \ket{1} \} \to \{\ket{-1},\ket{0}, \ket{1}, \ket{2} \}$. Now even and odd modes are sorted using MZ OAM sorter with $\alpha=\pi/2$\,(see Fig.\,\ref{mz1}). By passing the even modes alone through a DP and recombining the even and odd modes we finally obtain $\{\ket{-1},\ket{0}, \ket{1}, \ket{-2} \}$, which is the desired $X$ operation.

For the $X^2$ operation, we first sort the even and odd modes and perform transformation $\ket{\ell} \to \ket{\ell +2}$ only for the even OAM modes. This is followed by a reflection on all the modes and then we combine them on a  mode sorter. This results in $\{\ket{-2},\ket{-1}, \ket{0}, \ket{1} \} \to \{\ket{0},\ket{1}, \ket{-2}, \ket{-1} \}$.

Finally, for the $X^3$ operation, we first observe that $X^3= X^\dagger$. So we first sort even and odd modes and then pass the even modes through a DP to obtain $\{\ket{-2},\ket{-1}, \ket{0}, \ket{1} \} \to \{\ket{2},\ket{-1}, \ket{0}, \ket{1} \}$. Now by applying the transformation $\ket{\ell} \rightarrow \ket{\ell-1}$ using a spiral phase plate, which adds an OAM of $-\hbar$ per photon, we achieve the desired operation, namely, $\{\ket{2},\ket{-1}, \ket{0}, \ket{1} \} \to \{\ket{1},\ket{-2}, \ket{-1}, \ket{0} \}$.

Having introduced the necessary background, we present our device, polarization selective Dove prism, and derive the associated infinite dimensional operator in the next Section.

\section{Polarization selective Dove prism} \label{s3}

\begin{figure}
\centering
\includegraphics[scale=0.4]{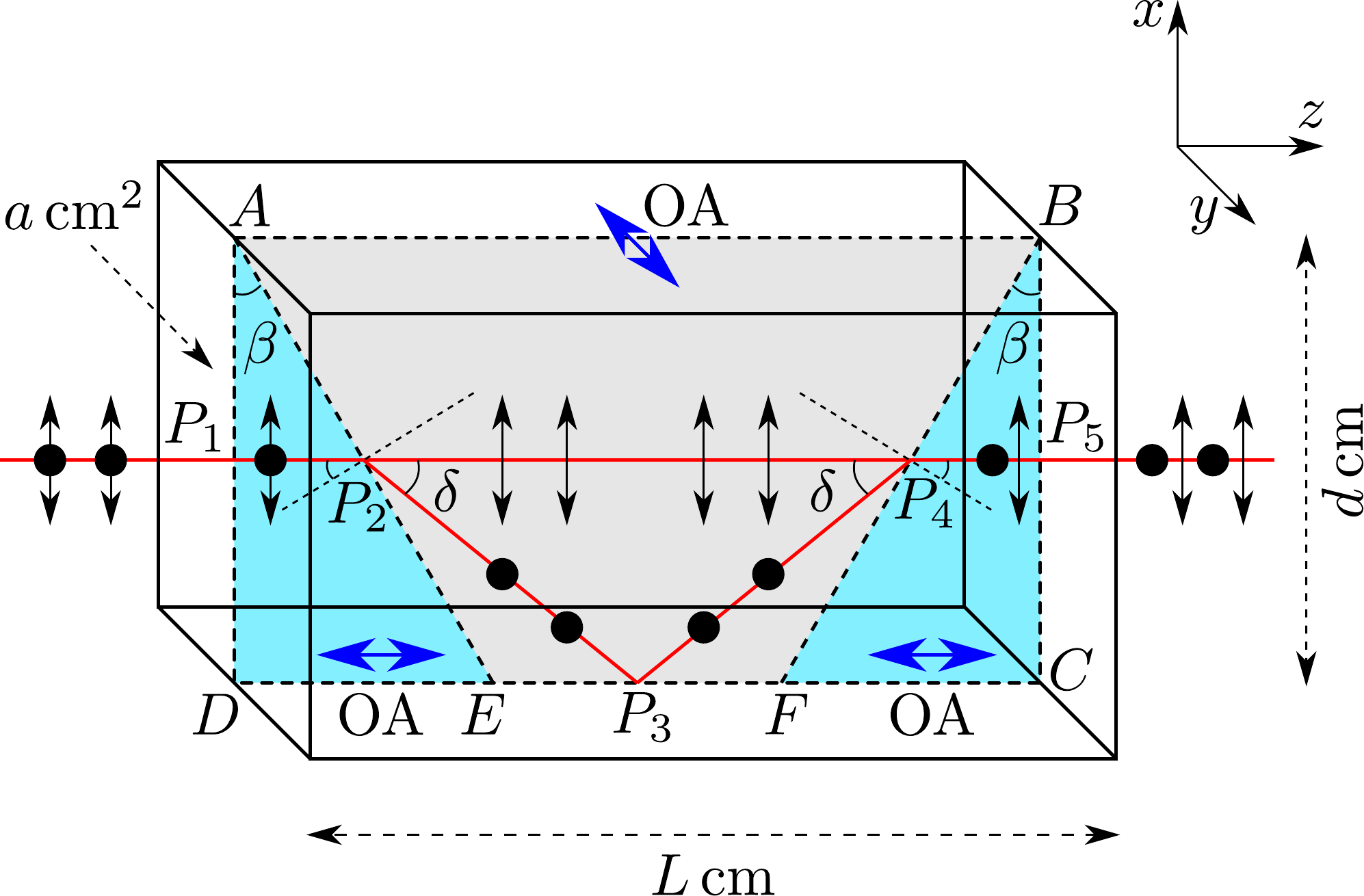}
\caption{Schematic diagram of the proposed device -- made of negative uniaxial crystal -- which acts as a DP for one of the orthogonal polarizations of the incoming light field. Optic axes\,(OA) of $AED$ and $BCF$ are along the $z$-axis, while that of $ABFE$ is along the $y$-axis. So both $x$\,(or the ordinary ray -- denoted by `$\updownarrow$') and $y$\,(or the extraordinary ray -- denoted by `$\bullet$') polarized light fields entering at point $P_1$ will ``see'' the same refractive index $n_o$ and thereby propagate undeviated. Now inside $ABFE$ the ordinary ray will pass through undeviated by experiencing a refractive index $n_o$. However, the extraordinary ray will experience a refractive index $n_e$, and therefore bends at $P_2$, total internally reflected at $P_3$, and reaches $P_4$. Finally, inside $BCF$ both ordinary and extraordinary rays experience the same refractive index, $n_o$, and hence propagate along the line $P_4P_5$. Distances and angles shown here are not to scale.}
\label{psdp1}
\end{figure}

\begin{figure}
\centering
\includegraphics[scale=0.55]{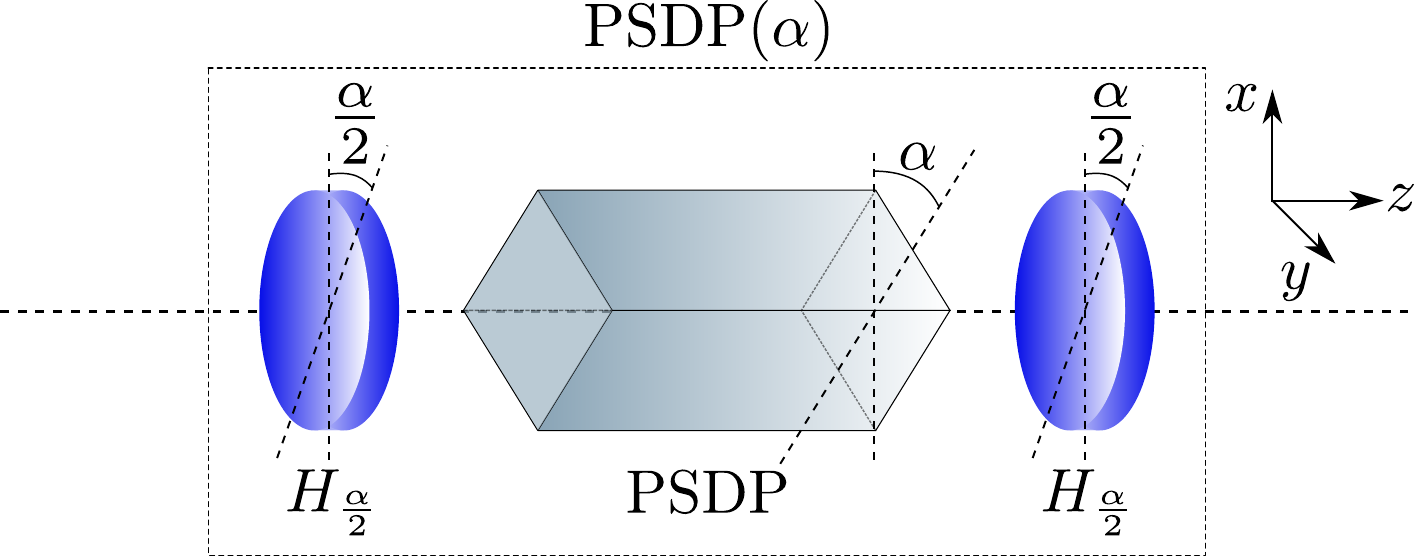}
\caption{Schematic diagram of the whole ${\rm PSDP}(\alpha)$ setup, which acts as a DP rotated through an angle $\alpha$ anticlockwise about the $z$-axis for the $y$-polarized light and leave the $x$-polarized light unaffected. This device consists of two half-wave plates\,(HWPs) rotated at an angle $\alpha/2$\,(denoted by $H_{\alpha/2}$) and a PSDP\,(Fig.\,\ref{psdp1}). Here, the Jones matrix corresponding to the PSDP is $J$\,(see Eq.\,(\ref{jon1})), while that corresponding to the entire optical setup is $-J_\alpha$\,(see Eq.\,(\ref{jon7})).}
\label{psdp2}
\end{figure}

Here we propose a device -- named PSDP -- which acts as a DP for one of the polarization states of the light without affecting the orthogonal polarization state. This device consists of a cube made of three negative uniaxial crystals glued together\,(see Fig.\,\ref{psdp1}). We use two half-wave plates to neutralize the effect of rotation of PSDP on the polarization states of light, as schematically outlined in Fig.~\ref{psdp2}. In this section, we describe PSDP and the role of each of its individual components in detail. 

 PSDP consists of three uniaxial crystals glued together as shown in Fig.~\ref{psdp1}. All three components of PSDP are constructed from  the same uniaxial crystal, however, their optic axis are aligned in different directions. The two components, at the beginning and at the end (represented by blue color), have their optic axis (OA) along $z$-axis and the middle symmetric trapezoidal component has the OA aligned along $y$-axis. All three components put together form a cuboid of length $L$ and square cross-section of   length $d$ and area $a = d^2$.

In a uniaxial crystal, the refractive index for ordinary ray is $n_o$ and for extraordinary ray it is $n_e$. Since the OA in the first and third component is along the direction of propagation of light, i.e., $z$-axis, both the polarization (along $x$- and $y$-axis) are ordinary rays for them. However, in the trapezoid the polarization along the OA is extraordinary, i.e., $y$-polarization and the polarization orthogonal to OA is ordinary, i.e., $x$-polarization.

Let a coherent paraxial light field of wavelength $\lambda$ propagating in the $z$-direction enter the cuboid from the left at $P_1$. Evidently, the light field propagates undeviated inside $AED$, as its OA is along the $z$-axis. At $P_2$, both polarizations ``see'' different refractive indices. As a result, the $y$-polarized light bends away from the normal and propagates towards $P_3$, whereas the $x$-polarized light propagates undeviated towards $P_4$. At $P_3$, the $y$-polarized light is totally internally reflected and propagates to $P_4$. Finally, both polarizations once again ``see'' same refractive index at $P_4$, as the OA of $BCF$ is along the $z$-axis and therefore propagate to $P_5$ undeviated.

The optical path traveled by the ordinary ray and the extra ordinary rays in a uniaxial crystal is different which results in delay in the arrival time for the two rays. This path difference can be compensated using a Soleil compensator\,(SC), or by choosing negative uniaxial crystals and appropriate angle $\beta$ as shown in Appendix\,\ref{opd}.

Mathematically, the operation corresponding to the PSDP can be written as
\begin{align} \label{jon1}
  J = \ket{x}\bra{x} \otimes \mathds{1} + \ket{y}\bra{y}\otimes \mathcal{U}_{\rm s}.
\end{align}
Here $\mathcal{U}_{\rm s}$ is DP operator acting on the OAM space\,[see Eq.\,(\ref{dov2a})], $\{\ket{x},\ket{y}\}$ are the $x$- and $y$-polarization states of light, and $\mathds{1}$ represents identity operator acting on the OAM space.

Now we consider a situation where the crystal is rotated through an angle $\alpha$ anticlockwise about the $z$-axis so that the plane of incidence  makes an angle $\alpha$ with the $x$-axis\,(Fig.\,\ref{psdp2}).  Rotating the PSDP setup by an angle $\alpha$ about the $z$-axis will result in the new operator given by 
\begin{align} \label{jon2}
  J(\alpha) = R_\alpha J_\alpha R_{-\alpha}, 
\end{align}
where $R_\alpha$ is the rotation matrix
\begin{align} \label{jon3}
R_\alpha = \begin{bmatrix}
\cos \alpha & -\sin \alpha \\
\sin \alpha & \cos \alpha
\end{bmatrix},
\end{align}
and 
\begin{align} \label{jon4}
J_\alpha = \ket{x}\bra{x} \otimes \mathds{1} + \ket{y}\bra{y}\otimes \mathcal{U}_{\rm s}(\alpha).
\end{align} 
For the rotated PSDP the states $R_\alpha\ket{x}$ is the ordinary ray and $R_\alpha\ket{y}$ is the extraordinary ray, and $\mathcal{U}_s(\alpha)$ is acting on the OAM states coupled with new extraordinary polarization.  However, we want $x$- and $y$-polarizations to remain the ordinary and extraordinary, i.e., the operator $\mathcal{U}_s(\alpha)$ should act on OAM states coupled with $y$-polarization.

To overcome this problem, we can employ two half-wave plates (HWP) rotated by angle $\alpha/2$, as shown in Fig.~\ref{psdp2}. The transformation matrix corresponding to a HWP is given by~\cite{simon89,hecht2017} 
\begin{align} \label{jon5}
H_0 = \begin{bmatrix}
\text{i} & 0 \\
0 & -\text{i}
\end{bmatrix}.
\end{align}
For the rotated HWP whose slow axis makes an angle $\alpha$ with the $x$--axis, the corresponding Jones matrix is
\begin{align} \label{jon6}
H_{\alpha/2} = R_{\alpha/2} H_0 R_{-\alpha/2} = \text{i}
\begin{bmatrix}
\cos \alpha & \sin \alpha \\
\sin \alpha & -\cos \alpha
\end{bmatrix}.
\end{align}
It can easily be seen that 
\begin{align} \label{jon7}
H_{\alpha/2} [R(\alpha) J_\alpha R(-\alpha)] H_{\alpha/2} &= 
H_0J_\alpha H_0 = -J_\alpha.
\end{align} 
Here, the negative sign introduces overall phase and can readily be discarded. Therefore, the action of the rotation on the uniaxial crystal can be compensated using two HWPs  and the whole setup result in the operator $J_\alpha$ which performs $\mathcal{U}_{\rm s}(\alpha)$ operation on the $y$-polarization and leave the OAM state associated with $x$-polarization unaffected. Hence, we achieve a device which is capable of performing polarization selective DP action.

In summary, PSDP is a passive device which makes use of in-principle lossless optical elements and can readily provide coupling between the OAM and the polarization DoFs. This desirable feature is highly sought after in fields such as optical quantum computation and quantum information. In the following Section we present a few applications of PSDP.

\section{Applications} \label{s4}
Although the PSDP is a all optical simple device, the applications of this device are limitless. It is not possible to present all the application in this article, as example, we show realization of CNOT and SWAP gates between polarization and OAM DoFs, non-interferometric OAM sorter and permutation operation in four OAM modes, which are some of the most important features required for quantum computation and communication.

\subsection{CNOT operation} \label{cn}
The action of the CNOT gate $C$ on a two-qubit system\,\cite{nielsen2010} is written as
\begin{align} \label{cn1}
  C\ket{i}\otimes\ket{j} = \ket{i}\otimes\ket{j\oplus i},
\end{align}
where $i,j \in \{0,1\}$ and $j\oplus i = j+i~{\rm mod}~2$. Since the state of the first qubit is not changing upon the action of $C$, it is called the control qubit and the second qubit is the target qubit. 

In our setup, the polarization can serve as the control qubit with the basis  $\{\ket{x},\ket{y}\}$ and OAM as the target system with the subspace spanned by $\{\ket{\pm\ell}\}$. The action of the PSDP with the operator $J(0)$ on the joint basis $\{\ket{x}\otimes \ket{\pm\ell},\ket{y}\otimes\ket{\pm\ell}\}$ reads
\begin{align}
  J(0):&\ket{x}\otimes\ket{\pm\ell} \to \ket{x}\otimes\ket{\pm \ell} \label{cn2} \\
  &\ket{y}\otimes\ket{\pm\ell} \to\ket{y}\otimes\ket{\mp \ell}, \label{cn3}
\end{align}
which is identical to the CNOT gate. Hence PSDP at $\alpha = 0$ realize the CNOT gate. 

\begin{figure}
\centering
\includegraphics[scale=0.3]{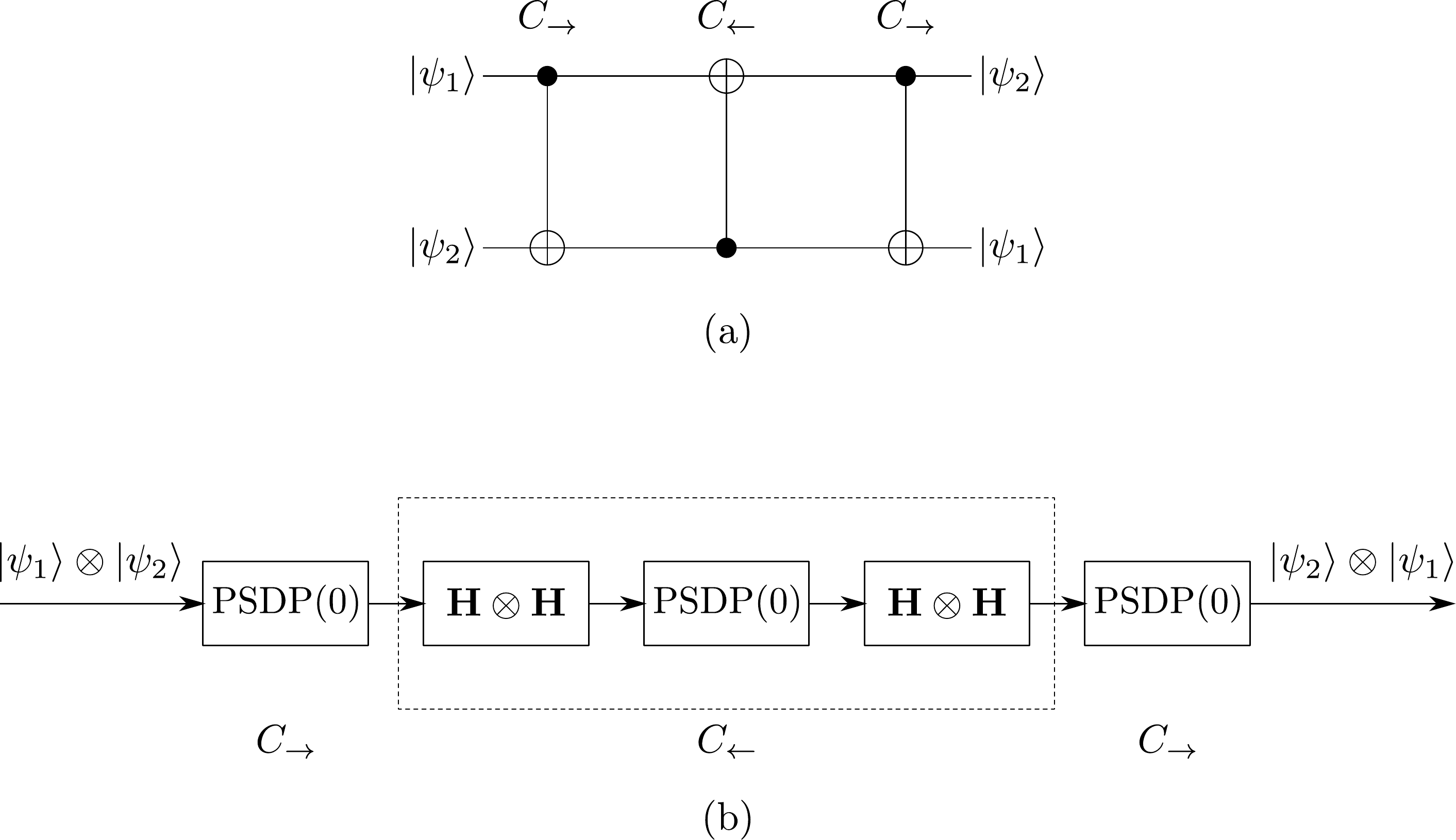}
\caption{(a) Circuit diagram of SWAP operation, implemented here using three CNOT gates. Here, $C_\rightarrow$\,($C_\leftarrow$) denotes that the first\,(second) qubit is the control qubit. (b) SWAP operation using three PSDP setups. ${\rm PSDP(0)}$ acts as a CNOT gate\,(denoted by $C_\rightarrow$; see (\ref{cn2}) and \ref{cn3}) with polarization qubit being the control qubit. Left and right multiplying $C_\rightarrow$ with $\mathbf{H} \otimes \mathbf{H}$, where $\mathbf{H}$ is the Hadamard operation\,[see Eq.\,(\ref{sw2})], we obtain $C_\leftarrow$ -- CNOT gate with OAM modes in basis $\{|\pm 1\rangle\}$ being the control qubit. }
\label{sw-fig}
\end{figure}

\subsection{SWAP operation} \label{sw}
SWAP operation is defined as follows~\cite{nielsen2010,escartin2013}:
\begin{align} \label{sw1}
  \text{SWAP}: \ket{\psi_1}\otimes\ket{\psi_2} \to \ket{\psi_2}\otimes\ket{\psi_1}.
\end{align}
SWAP gate on two-qubit system can be decomposed into three CNOT operation, i.e.,~(Fig.~\ref{sw-fig})
\begin{align} \label{sw1a}
  \text{SWAP} = C_\rightarrow C_\leftarrow C_\rightarrow,
\end{align}
where $C_\rightarrow$\,($C_\leftarrow$) is the CNOT operation in which the first qubit is the control\,(target) qubit and the second one is the target\,(control) qubit.

SWAP gate is an important gate for quantum computation and quantum information. In general it is a difficult operation to implement,  especially in an optical system. However, the SWAP operation can be implemented using a set of PSDP as follows: as shown in the subsection\,\ref{cn}, ${\rm PSDP}(0)$ acts as a CNOT operation (say $C_\rightarrow$) on polarization plus OAM DoFs where $\{\ket{x},\ket{y}\}$ is the basis for polarization and $\{\ket{\pm 1}\}$ for OAM, and polarization is the control qubit. The operation $C_\rightarrow$ can be converted into $C_\leftarrow$ by using Hadamard operations on the two qubits, i.e.,  $C_{\leftarrow}=(\mathbf{H} \otimes \mathbf{H}) C_\rightarrow (\mathbf{H} \otimes \mathbf{H})$, where $\mathbf{H}$ is the Hadamard operation given by~\cite{nielsen2010}
\begin{align} \label{sw2}
\mathbf{H} =\frac{1}{\sqrt{2}} 
\begin{bmatrix}
1 & 1 \\
1 & -1
\end{bmatrix}.
\end{align}
It can be observed that the Hadamard operation on the polarization qubits can be realized through the use of rotated HWP, $H_{\pi/8}$, whereas the same on the OAM mode qubit with basis $\{|\pm 1 \rangle\}$ is realized using $\pi/2$- and $\pi$-phase converters which can be realized using thin cylindrical lenses of positive focal length~\cite{padgett99b}. With this,  the SWAP gate on polarization and OAM DoFs can be implemented using three PSDP setups. The explicit construction of SWAP gate is given in Fig.~\ref{sw-fig}.

\subsection{OAM sorting using PSDP} \label{osp}

\begin{figure}
\centering
\includegraphics[scale=0.4]{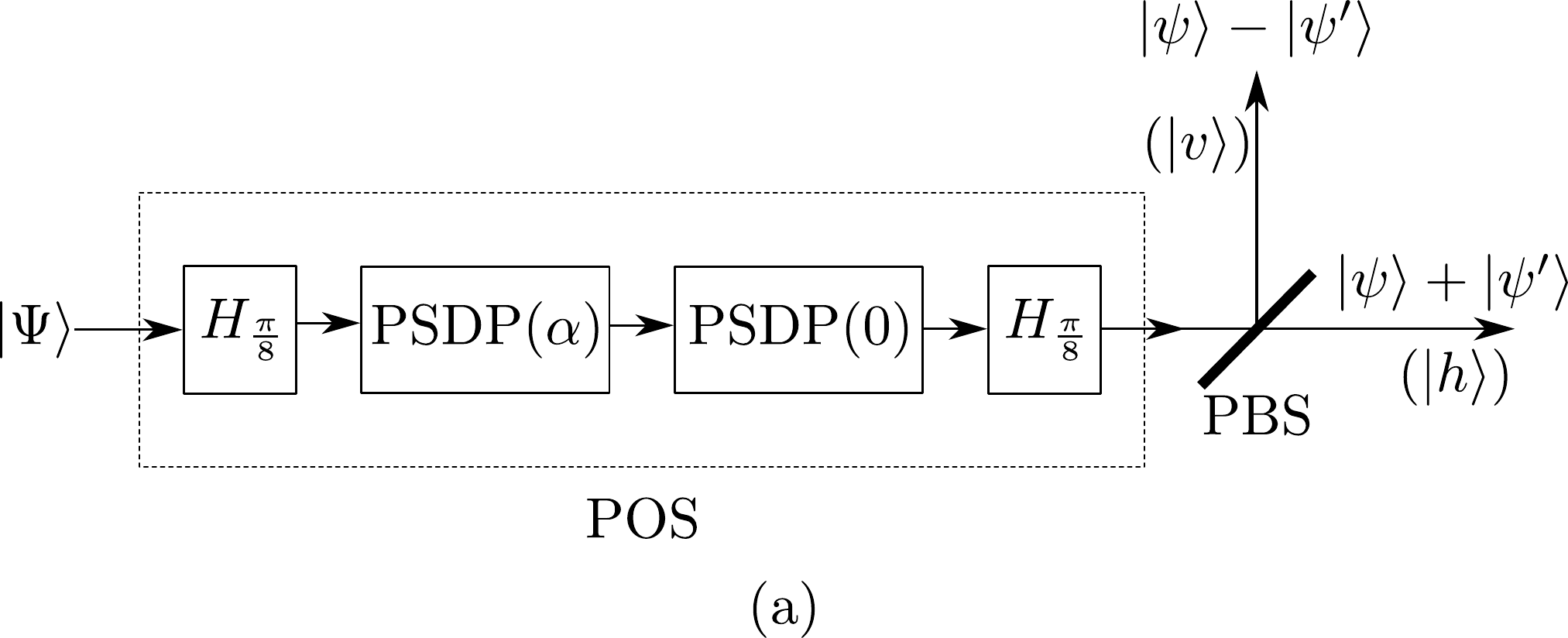} \\
\includegraphics[scale=0.3]{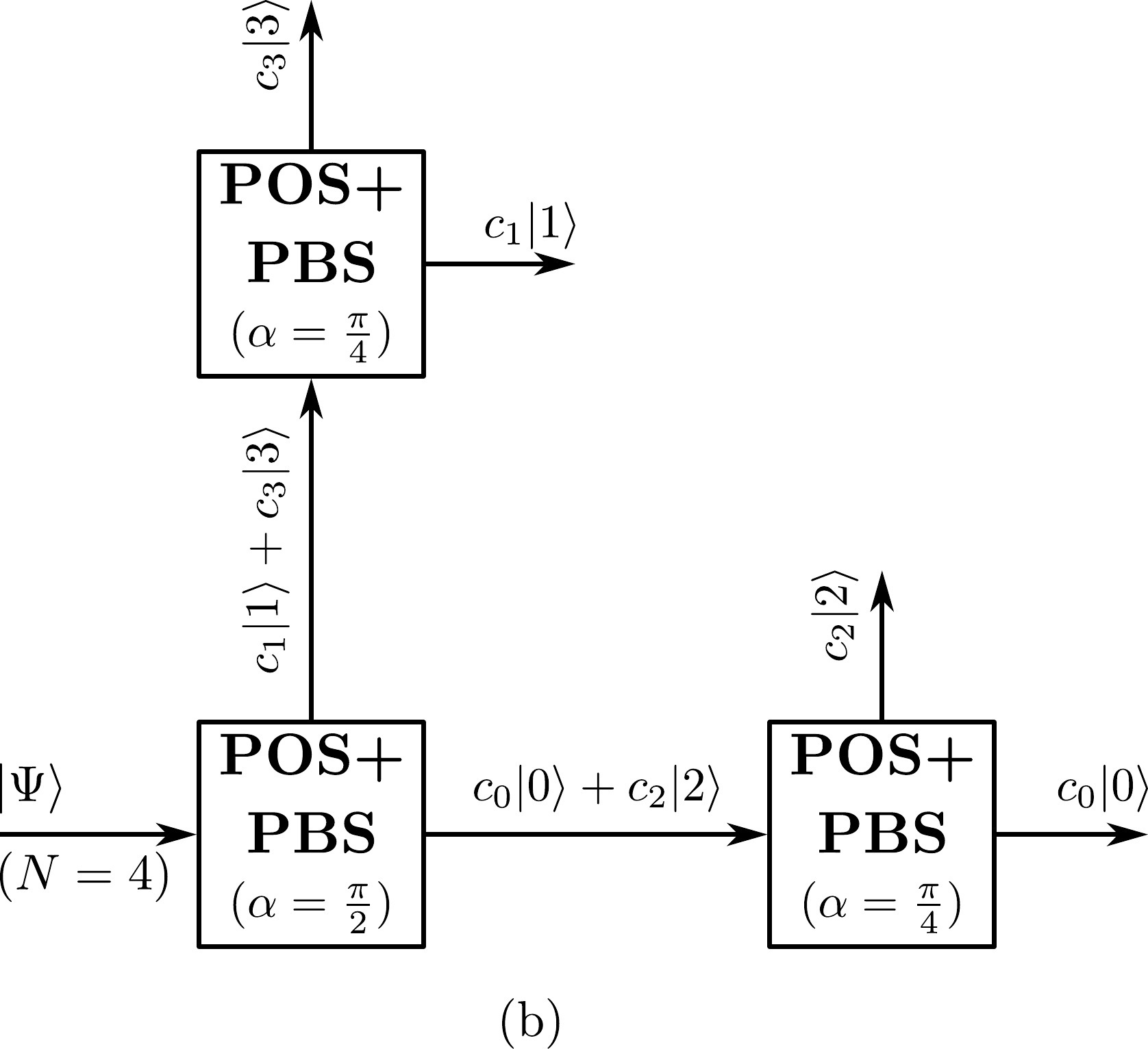}
\caption{(a) Schematics diagram for PSDP OAM sorter\,(POS) which sorts $N$ LG modes in the polarization domain. This device consists of two rotated HWPs\,($H_{\frac{\pi}{8}}$) and two ${\rm PSDP}(\alpha)$\,[with rotation angles $\alpha$ and 0\,(Fig.\,\ref{psdp2})]. If the incoming light is in the state $|\psi \rangle\otimes\ket{h}$, the state of light after passing through POS will be $(|\psi \rangle+|\psi' \rangle)/\sqrt{2}\otimes\ket{h} +(|\psi \rangle-|\psi' \rangle)/\sqrt{2}\otimes\ket{v}$. Polarizing beamsplitter\,(PBS) is used to split the resultant light into horizontal and vertical beams conditioned on the polarization states.  (b) OAM sorting of four OAM modes using three POSs. }
\label{pos}
\end{figure}

We now demonstrate how PSDP can sort the constituent LG modes present in a paraxial light field. First we present the scheme to sort even-odd OAM modes. The setup for this is given in Fig.~\ref{pos}(a). This setup requires two HWPs, two PSDPs and a polarizing beamsplitter.

Consider a paraxial light field in horizontal polarization represented by the state vector 
\begin{align} \label{s4e1}
|\Psi \rangle &= \sum_\ell c_\ell \ket{\ell} \otimes |h \rangle, 
\end{align}
where $c_\ell$ are complex coefficients. In our setup, the beam passes through the HWP rotated at $\pi/8$ angle $H_{\pi/8}$\,[see (\ref{jon6})] which causes the transformation
\begin{align} \label{s4e2}
|\Psi \rangle \rightarrow \sum_\ell c_\ell \ket{\ell} \otimes\frac{1}{\sqrt{2}}\,  (|h \rangle + |v \rangle) \equiv |\Psi_1 \rangle.
\end{align}
The action of PSDP$(\alpha)$ followed by another PSDP$(0)$ transform the state $\ket{\Psi_1}$ to 
\begin{align} \label{s4e3}
  \ket{\Psi_2}  =   \frac{1}{\sqrt{2}}\sum_\ell  c_\ell\left( \ket{\ell}\otimes|h \rangle + e^{\text{i}2\ell\alpha} \ket{\ell}\otimes|v \rangle \right). 
\end{align}

Finally, the action of another $H_{\pi/8}$ results in
\begin{align} \label{s4e4}
|\Psi_3 \rangle &=\frac{1}{2}\sum_\ell  c_\ell \left[ (e^{\text{i}2\ell\alpha} +1)\ket{\ell}\otimes|h \rangle \right. \nonumber \\ 
  &\,\,\, \left. + (1-e^{\text{i}2\ell\alpha}) \ket{\ell}\otimes|v \rangle \right].
\end{align} 
Therefore, choosing $\alpha = \pi/2$ will result in a state entangled between OAM and polarization DoFs, which is given by
\begin{align} \label{s4e5}
|\Psi_3 \rangle &=\sum_{n=-\infty}^\infty  c_{2n}\left(\ket{2n}\otimes|h \rangle + \ket{2n+1}\otimes|v \rangle\right).
\end{align}
Here all the even OAM modes are coupled with the horizontal polarization and all the odd modes are coupled with vertical polarization. The even and odd modes can now be separated in different spatial modes using a polarizing beamsplitter. 

The even (and odd) modes can further be separated into $4n$ and $4n+2$ modes by choosing $\alpha = \pi/4$. Hence, recursive use of the PSDP based OAM sorter can be employed to sort all the OAM modes. In order to sort $N$ OAM modes we require $N-1$ PSDP OAM sorter (POS), which include $2N-2$ HWPs, $2N-2$ PSDP setups and $N-1$ polarizing beamsplitters. However, this setup does not require aligning interferometers; hence, PSDP can sort OAM states efficiently and the setup is scalable. In Fig.~\ref{pos}(b) we sketch the sorting of four OAM modes using three POS.

\subsection{Generalized $X$ operation}

Generalized Pauli $X$ operation is implemented on the OAM subspace spanned by $\{|-2 \rangle, |-1 \rangle, |0 \rangle, |1 \rangle\}$ using PSDPs as given below. Without loss of generality we can assume that the input light is in the horizontal polarization state. First step in implementing $X$ operation is to add unit angular momentum. This can be done using a spiral phase plate. This results in the transformation $\{\ket{-2},\ket{-1}, \ket{0}, \ket{1} \} \to \{\ket{-1},\ket{0}, \ket{1}, \ket{2} \}$. Now letting the resultant modes pass through POS\,(see Fig.~\ref{pos}\,(a)) with $\alpha=\pi/2$ changes the polarization state of the even modes, $\ket{0}$ and $\ket{2}$, from $\ket{h}$ to $\ket{v}$. However, the polarization state of the odd modes, $\ket{-1}$ and $\ket{1}$, remains the same. Subsequent action of ${\rm PSDP}(0)$ will result in the even modes transforming as $\{\ket{0},\ket{2} \} \to \{\ket{0},\ket{-2}\}$, while the odd modes\,(in $\ket{h}$) remain unaffected. Finally, passing these modes through ${\rm POS}^{-1}$ -- obtained by interchanging ${\rm PSDP}(\alpha)$ and ${\rm PSDP}(0)$ in ${\rm POS}$\,(see Figure\,\ref{pos}\,(a)) with $\alpha=\pi/2$ -- we obtain $\{|-1 \rangle, |0 \rangle, |1 \rangle, |-2 \rangle\}$ (all available now in $\ket{h}$).  Since $X^3 = X^\dagger$, the same setup in the reverse order can be used to implement $X^3$ operation.

In the case of $X^2$ operation, we first pass the given spatial modes through POS with $\alpha=\pi/2$ so that the polarization state of the even modes is changed from $\ket{h}$ to $\ket{v}$. Next step in realizing $X^2$ operation is to add two units of OAM in only the even states. This can be achieved by using fork hologram in spatial light modulators (SLM). SLMs can be made to operate only on a certain polarization state leaving the other polarization unaffected. We can use a DP to perform $\ket{\ell} \to\ket{-\ell}$ transformation for all the modes.  Finally, by applying ${\rm POS}^{-1}$ transformation with $\alpha=\pi/2$ results in $X^2$ operation. Hence, using two POS setups and one PSDP setup one can realize generalized Pauli $X$ and $X^3$ operations and using SLM and DP along with two POS setups $X^2$ operation, all in a single beam.

\section{Conclusion} \label{s6}
We have proposed an all optical device named polarization sensitive Dove prism which can couple OAM and polarization DoFs of light without involving digital/liquid crystal devices or interferometric methods. 
This device consists of lossless linear optical elements such as uniaxial crystal and half wave-plates which makes the device highly efficient.  As applications of PSDP, we present the scheme to implement CNOT operation and SWAP operation between polarization and OAM of light, efficient and scalable sorting of OAM states and implementation scheme for cyclic permutation operations on a four-dimensional  subspace of OAM of light. Since, OAM is a highly sought after property for quantum information processing tasks, PSDP can be an important device for photonic quantum computation and information.

\begin{acknowledgments}

Yasir acknowledges funding from the SERB-NPDF scheme of Government of India\,(File No. PDF/2019/001881). S.K.G. acknowledges the financial support from SERB-DST (File No. ECR/2017/002404).
\end{acknowledgments}

\appendix

\section{Optical path difference (OPD)} \label{opd}

In this appendix we calculate the optical path difference\,(OPD) between the ordinary and extraordinary rays as shown in Fig.\,\ref{psdp1}. We have
\begin{align} \label{eq1}
{\rm OPD} &= 2n_e P_2P_3 - 2n_o P_2P_3 \cos \delta \nonumber \\ 
&= 2 P_2 P_3 (n_e - n_o \cos \delta).
\end{align}
Because $P_2P_3 \sin \delta = d/2$, the OPD is
\begin{align} \label{eq2}
{\rm OPD} = \frac{d\,(n_e - n_o \cos \delta)}{\sin \delta}.
\end{align}
Relation between angles $\delta$ and $\beta$ can be found with the aid of Snell's law at $P_2$ for the extraordinary ray. We find that
\begin{align} \label{eq3}
n_o \sin \beta &= n_e \sin (\beta+\delta) \,\,\,{\rm (or)} \nonumber \\
\tan \beta &= \frac{\sin \delta}{(n_o/n_e)-\cos \delta}.
\end{align}
This OPD between the ordinary and extraordinary rays can be compensated by using an SC~\cite{hecht2017,iizuka2002}. Evidently, the OPD in Eq.\,(\ref{eq2}) can be made 0 for the choice
\begin{align} \label{eq4}
\delta = \cos^{-1} \left( \frac{n_e}{n_o} \right).
\end{align}
Then by (\ref{eq3}) the corresponding $\beta$ is
\begin{align} \label{eq5}
\beta = \tan^{-1} \left( \frac{n_e}{\sqrt{n_o^2-n_e^2}} \right).
\end{align} 
This choice of $\beta$ ensures that the OPD is 0, and we don't require an SC. However, we must ensure that the extraordinary ray at point $P_3$ in Fig.\,\ref{psdp1} is total internal reflected for the choice of $\delta$. For example, if we consider calcite\,($n_e=1.486$ and $n_o=1.658$ when $\lambda=589.3\,{\rm nm}$ is used~\cite{ghatak2010}), a negative uniaxial crystal, the angles $\delta$ and $\beta$ are $26.329^\circ$ and $63.671^\circ$, respectively.


\twocolumngrid
    
\bibliography{refer}

\end{document}